\documentclass[a4paper, twocolumn, superscriptaddress,prl]{revtex4-1}  
\usepackage{graphicx}
\usepackage{amsmath,amssymb}

\newcommand{\ket}{\bigr \rangle}
\newcommand{\bra}{\bigl\langle}
\usepackage{epstopdf}

\begin{document}
\title{Quantum superresolution in fluorescence microscopy}
\author{O. Schwartz}
\author{J.M. Levitt}
\author{R. Tenne}
\author{S. Itzhakov}
\author{Z. Deutsch}
\author{D. Oron}
\affiliation{Department of Physics of Complex Systems, Weizmann Institute of
Science, Rehovot 76100, Israel}

\title{Quantum Superresolution in Fluorescence Microscopy}
\begin{abstract}The optical diffraction limit, formulated by Abbe 140 years ago, imposes a bound on imaging resolution in classical optics. Over the last twenty years, many theoretical schemes have been presented for overcoming the diffraction barrier in optical imaging using quantum properties of light. An experimental realization of sub-diffraction limited quantum imaging has, however, remained elusive. Here, we demonstrate a quantum imaging method taking advantage of non-classical light naturally produced in fluorescence microscopy due to photon antibunching, a fundamentally quantum phenomenon prohibiting simultaneous emission of multiple photons. Using a photon counting digital camera, we detect antibunching-induced second and third order intensity correlations and perform sub-diffraction limited quantum imaging in a standard wide-field fluorescence microscope.
\end{abstract}


\maketitle

The optical diffraction limit restricts the resolution of far-field optical microscopes to approximately half the wavelength of light. Abbe's description of the imaging system\cite{Abbe1873} is based on the laws of classical linear optics, applied to stationary objects. Correspondingly, there are three loopholes in the argument, concerning the linearity, stationarity, and classicality assumptions. In the last two decades, super-resolution imaging methods were developed based on nonlinear optical effects such as stimulated emission\cite{STED_Hell_OL1994}, optical shelving\cite{Hell_RESOLFT_2004} and fluorescence saturation\cite{Gustafsson_SSI_PNAS05}. More recently, sub-diffraction limited imaging was achieved by another class of microscopy methods, making use of non-stationary emission of fluorescent markers caused either by photo-switching\cite{PALM_Betzig_Science2006,
STORM_NatMet06}, or by intrinsic brightness fluctuations \cite{Localization_by_blinking_Heintzmann_OpEx2005, SOFI_DertingerPNAS2009}.

The remaining loophole for overcoming the diffraction barrier, resorting to quantum optics, has received a lot of attention in the recent years. It has been shown that high order quantum interference patterns arising in quantum optics can yield spatial distribution of correlations much tighter than classically allowed. Such fringes have been observed using coincidence detection in various settings \cite{Four_photon_Interference_Zeilinger_Nature2004, Afek_Science2010, Boyd_CentroidsPRL2011}. It seems tempting to use these sharp spatial features to image subwavelength details of microscopic objects. Several quantum superresolution schemes have been proposed, utilizing multi-mode squeezed light \cite{Kolobov_Quantum_Limits_PRL2000}, an arrangement of single photon emitters \cite{Zanthier_Quantum_imaging_Incoherent_PRL2007}, or generalized quantum states of light\cite{saleh2005wolf, Centroids_PRL2009, Shapiro_Sub_Rayleigh_PRA2009}.%
At the same time, although quantum optical methods have enabled image entanglement\cite{Entangled_Images_Science2008} and sub-shot noise imaging\cite{Sub_shot_noise_imaging_NatPhot2010}, as well as quantum optical coherence tomography with improved depth resolution\cite{QuantumOCT_PRL2003}, sub-diffraction limited quantum imaging has not yet been experimentally demonstrated.

The common element of most proposed quantum superresolution schemes is illuminating an absorptive sample with a nonclassical state of light. An alternative approach, proposed theoretically by Hell et al.\cite{Hell_photon_pairs1995Bioimaging}, relies on non-classical properties of light emitted by the sample itself, while using regular laser light for illumination. It was shown in this work that a hypothetical quantum emitter producing photons only in pairs (or groups) can be imaged using coincidence detection with a resolution increase similar to that attainable in two-photon (multiphoton) microscopy. Unfortunately, the technique has not been taken up experimentally since no suitable multiphoton emitting fluorophore has ever been introduced.

\begin{figure*}
  \includegraphics[bb=10 0 864 280, scale=0.8]{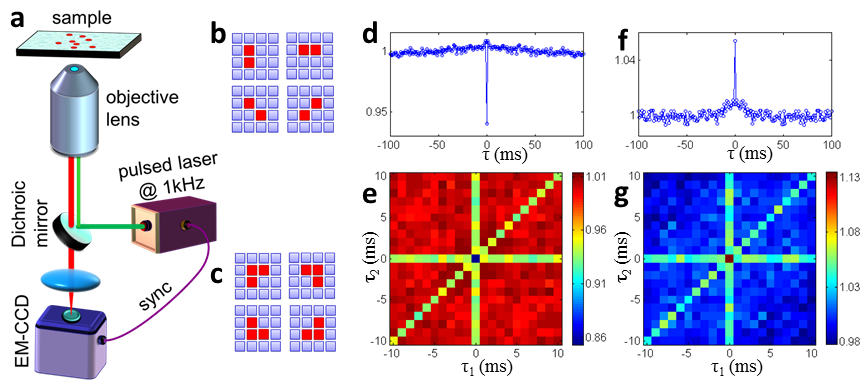}\\
  \caption{\textbf{Detecting nonclassical correlations in fluorescence microscopy.} \textbf{a},  A schematic of the experimental setup used for correlation imaging. \textbf{b} and \textbf{c}, The pixel configurations used for second and third order correlation detection, respectively. \textbf{d}, a typical second order intensity autocorrelation curve from an eight-pixel region on the CCD covering an image of a single emitter, computed as $g^{(2)}(\tau)=\sum_{i\neq k} \bra n_i(t) n_k(t+\tau) \ket$, where the sum runs over all pixel pairs in the relevant region. The graph features an antibunching dip. (e) demonstrates a two-dimensional plot representing third order autocorrelation $g^{(3)}(\tau_1,\tau_2)=\sum_{i\neq k} \bra n_i(t) n_k(t+\tau_1) n_m(t+\tau_2) \ket$ , where the sum runs over all three-pixel combinations within the same region of interest. \textbf{f}, a correlation function computed for the same region on the CCD illuminated with a classical light source with a bunching peak at zero delay. \textbf{g}, the third order correlation   computed with classical signal demonstrates behavior opposite to antibunching: it has ridges at $\tau_1=0$ , at $\tau_2=0$  and at $\tau_1=\tau_2$ , and a peak at $\tau_1=\tau_2=0$ .}\label{setup}
\end{figure*}

Here, we extend the idea of multi-photon detection put forward by Hell et al.\cite{Hell_photon_pairs1995Bioimaging} by utilizing fluorophores in which emission of more than one photon is suppressed. This phenomenon, known as photon antibunching\cite{Antibunching_Kimble_PRL1977}, is observed in most common fluorophores, such as organic dyes \cite{Antibunching_dye_Basche_PRL1992} or quantum dots\cite{Antibunching_Alivisatos_CPL2000,Antibunching_QDs_Buratto_Nature2000}, even under ambient conditions. The necessary quantum emitters are thus widely used in fluorescence microscopy, a ubiquitous life science imaging tool. Due to photon antibunching, in every point of the image plane of a fluorescence microscope photon statistics is sub-Poissonian \cite{Sub-Poissonian_Mandel_OL1979}, i.e. the number of simultaneous multi-photon detection events is smaller in every order than it is for classical light. Quantifying the missing N-photon coincidence events gives a signal equivalent to N-photon detection signal, narrowing the effective point spread function by a factor of $\sqrt{N}$ \cite{schwartz2012improved}. In this work, we detect photon statistics in the image plane of a wide-field fluorescence microscope, determine the spatial distribution of missing two- and three-photon coincidence events, and reconstruct second and third order superresolved images.

Our measurement scheme relies on detecting fluorescence intensity correlations. When a perfectly antibunched quantum emitter is imaged onto a detector array, the single photon it emits following excitation can arrive to only one of the detectors. While classically the detector readings would be uncorrelated, antibunching creates negative correlations between them. We can thus define the second order antibunching signal as the same-time two-point correlation function $A^{2}(x,y)=\bra I(x) I(y) \ket - \bra I(x)\ket\bra I(y) \ket$, where  $I(x)$ and  $I(y)$ are fluorescence intensities at points $x$ and $y$, and the angle brackets denote averaging over time. Higher order antibunching signals can be similarly defined as N-point irreducible intensity correlation functions. Importantly, contributions to the antibunching signal from individual fluorophores are additive, and can therefore be utilized as a local measure of the emitter density, thus directly providing superresolved images of the fluorophore spatial distribution.
\begin{figure}
  \includegraphics[bb=40 0 350 450,width=90mm]{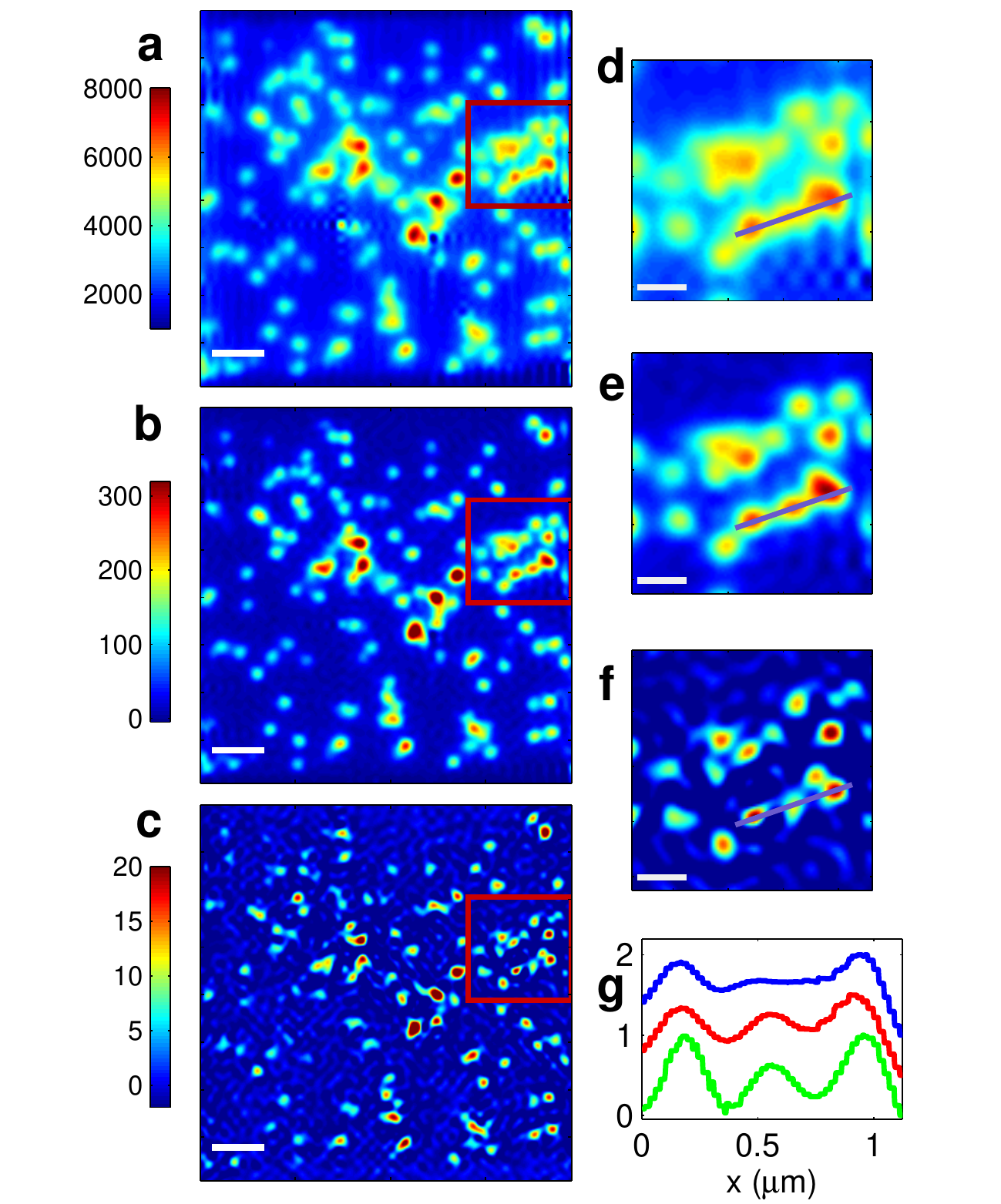}\\
  \caption{\textbf{Fluorescence antibunching imaging.} Panels \textbf{a-c} show a regular fluorescence image, second and third order antibunching images, respectively. Scalebar length is $1\mu$m. \textbf{d-f}, magnified views of the boxed areas in the images \textbf{a-c}. Scalebar is $400$nm. The graph presented in panel \textbf{g} shows a cut of the magnified images (in normalized units) along the line indicated in panels \textbf{d-f}. The blue line corresponds to regular fluorescence image, the red and green lines represent the second and third order antibunching imaging, respectively. The lines are vertically shifted for visibility.   }\label{defocusing}
\end{figure}

For the proof of principle demonstration of antibunching imaging, we used test samples consisting of CdSe/CdS/ZnS colloidal quantum dots (QDs), commonly used as labels in fluorescence microscopy\cite{QDs_Michalet_Weiss_Science2005}, with the main emission peak at 617 nm. The QDs were embedded in a thin polymer film spin-coated onto a glass slide (see Supplementary Information for further details). Such QDs demonstrate strong photon antibunching due to exciton-exciton annihilation interactions inhibiting emission of more than one photon following photoexcitation\cite{Antibunching_Alivisatos_CPL2000}. These QDs were recently shown to exhibit nearly absolute photostability under similar conditions \cite{schwartz2012colloidal}.

Experimental observation of nonclassical intensity correlations was carried out using a regular wide-field epifluorescence microscope shown schematically in Fig.~1a. To detect photon statistics simultaneously in the entire field of view, we used an electron multiplying charge-coupled device (EM-CCD) in the photon counting mode as a fluorescence detector. The fluorophores were excited with 300 ps laser pulses at 532 nm, with pulse energy close to saturation. Following emitter relaxation, the image was read out and stored in a computer. The excitation pulse/image readout sequence was repeated at a rate of 1 kHz. The pixel readings were thresholded to produce maps of photon detection events, neglecting the probability of detecting more than one photon per pixel in the same excitation cycle.

Pulsed excitation in combination with image readout after every exposure allowed us to detect the second order (temporal) intensity correlation functions $g_{ik}^{(2)}(\tau)=\bra n_i(t) n_k(t+\tau)\ket$ , where $n_i$  and  $n_k$ are the numbers of photons detected in pixels $i$ and $k$  in frames $t$ and $t+\tau$, respectively, and the angle brackets denote averaging over t. A typical correlation function shown in Fig.~1d exhibits the characteristic antibunching dip at zero delay. Furthermore, we were able to detect antibunching features in the third order intensity autocorrelation function $g_{ikm}^{(2)}(\tau_1,\tau_2)=\bra n_i(t) n_k(t+\tau_1) n_k(t+\tau_2)\ket$, depending on two discrete delay times  $\tau_1$ and $\tau_2$. A two-dimensional plot of a typical third order temporal correlation function is shown in Fig.~1e. The plot features depressed lines at $\tau_1=0$, at $\tau_2=0$  and at $\tau_1=\tau_2$, which represent the lack of two-photon coincidence events. The central data point of this plot, at $\tau_1=\tau_2=0$, which is depressed even further, corresponds to the missing three-photon coincidence events. The dips in the temporal autocorrelation functions represent the non-classical signal that we proceed to utilize to produce superresolved images. The observed magnitude of the antibunching features is reduced due to frame-to-frame fluctuations in the CCD readout circuitry, leading to apparent bunching of detection events. Our data processing offsets the antibunching signal to account for this effect.
To quantify the second order quantum correlations at every point in the image plane with sub-diffraction limited resolution, we computed the cross-correlations between pairs of neighboring pixels in configurations shown in Fig.~1b. The resulting four correlation maps were Fourier-interpolated and summed. Similarly, the third order antibunching images were obtained by computing the third order cross-correlation for pixel configurations shown in Fig.~1c. The details of the data processing are described in the Supplementary Information.
A typical superresolved image set is shown in Fig.2. A regular fluorescence image and the second and third order antibunching images of the same area, presented in Fig.~2a-c, demonstrate consecutive improvement of resolution. The magnified view of a small region (Fig.~2d-f) illustrates the initially unresolved features of QD distribution revealed by antibunching imaging. Enhanced resolution is also evident in the line scan plotted in Fig.~2g. Quantitatively, the resolution defined as the full width at half maximum of the point spread function improves from 272 nm in regular images to 216 nm in the second order and 181 nm in the third order, corresponding to a resolution enhancement by a factor of 1.5 (see Supplementary Information for resolution quantification and additional antibunching images).

\begin{figure}
  \includegraphics[width=90mm]{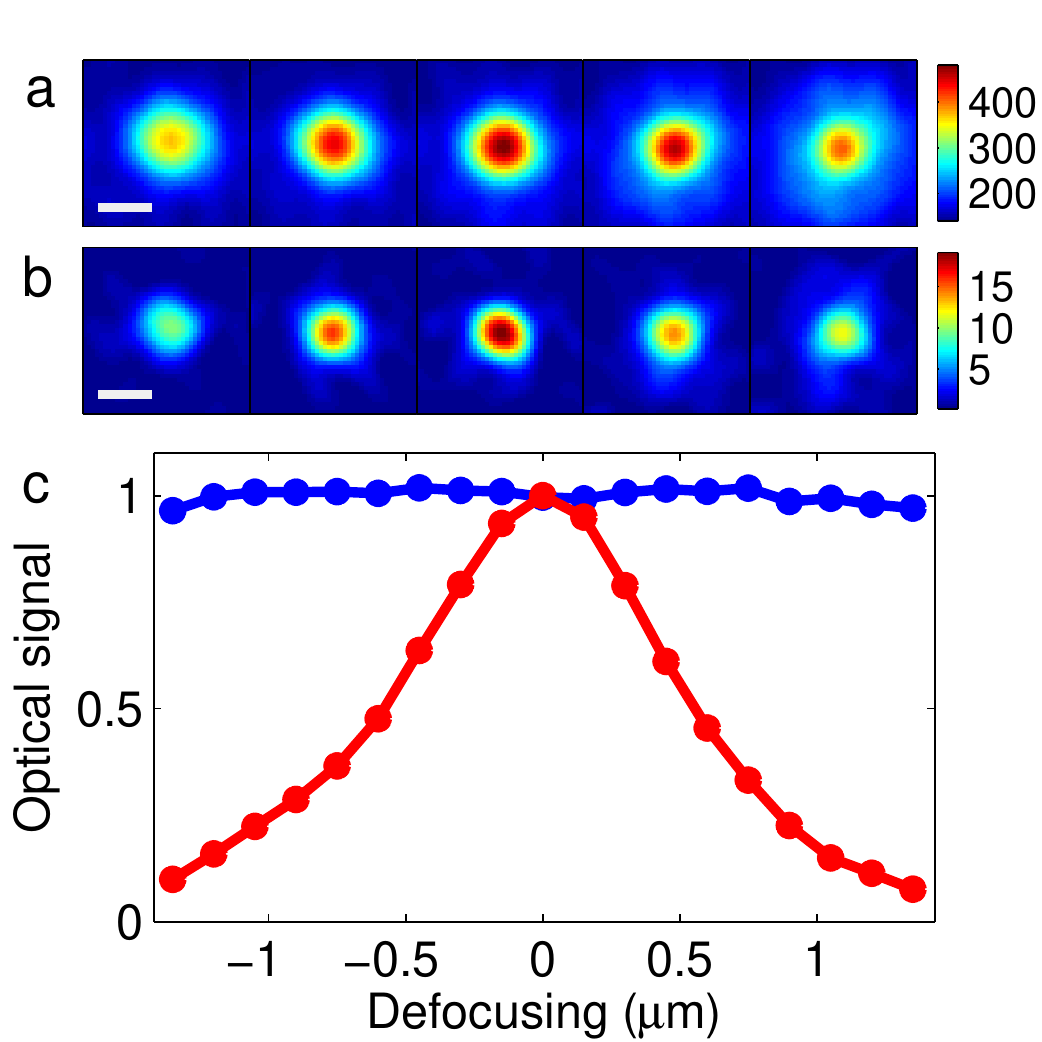}\\
  \caption{\textbf{Regular fluorescence and second order antibunching imaging dependence on defocusing.} Panels \textbf{a} and \textbf{b} show, respectively, regular fluorescence and 2nd order antibunching images of a QD with the imaging system focused (left to right) at -880 nm, -440 nm, 0, 440nm, 740 nm relative to the sample. Scale bar is 250 nm. Panel \textbf{c} shows the total signal integrated over the field of view, as a function of defocusing. The blue line represents regular fluorescence. The red line corresponds to second order antibunching imaging. The integrated fluorescence signal is practically constant as a function of depth. In contrast, the integrated antibunching signal decays quickly with defocusing. }\label{defocusing}
\end{figure}

In addition to improving the transverse resolution, antibunching imaging has an optical sectioning capability similar to that demonstrated by multi-photon excitation microscopy. This is illustrated by the out-of-focus images of a QD presented in Fig.~3. A comparison of the regular fluorescent images of Fig.~3a with the corresponding second order antibunching images shown in Fig.~3b demonstrates that the antibunching signal decreases faster with defocusing than regular fluorescence. While the regular fluorescence signal is blurred by defocusing, in transparent medium its integral remains unchanged, leading to out-of-focus background in wide-field microscopy. In contrast, the second order antibunching signal fades away with defocusing, leading to discrimination of the out-of-focus signal. Fig.~3c shows variation of the total signal, integrated over the field of view, as a function of defocusing for regular fluorescence imaging and for the second order antibunching signal.

The antibunching signal is a quantum optical observable quantifying the dip at zero delay in the intensity correlation functions, which has no analog in classical optics. Equivalently, it can be thought of as originating from sub-Poissonian photon statistics of antibunched light, which cannot be interpreted in terms of classical electrodynamics either\cite{Reduced_Fluctuations_Zoller_PRL1981}.

The antibunching based superresolution imaging method presented here is closely related to Superresolution Optical Fluctuation Imaging (SOFI) \cite{SOFI_DertingerPNAS2009}. In both approaches, intensity correlations are detected in the image plane of a fluorescence microscope and used to produce superresolved images. The main difference between the two is the source of the signal fluctuations recovered via correlations: fluorophore brightness fluctuations leading to super-Poissonian photon statistics in SOFI, and antibunched emission resulting in sub-Poissonian statistics in case of antibunching microscopy. Correspondingly, the SOFI signal is highly dependent on the character of emission fluctuations exhibited by the fluorophores, which varies widely from one species to another\cite{Power_law_intermittency_Orrit2007}, while in antibunching imaging the statistical properties of signal giving rise to the superresolved images are universal, arising from steady emission of fluorophores with no fluctuations other than antibunching-modified shot noise.

Performance of antibunching imaging is limited at present by the parameters of the camera used for photon detection. The rapid progress of the photon detector technologies in the recent years gives hope that fast and low-noise detectors will become available in the near future, which will dramatically improve the practical prospects of antibunching imaging.

In summary, we have demonstrated a superresolution imaging technique enabled by the quantum properties of light inherent in fluorescence microscopy, bringing quantum imaging a step closer to practical applications.

Financial support by the European Research Council Starting Investigator Grant SINSLIM 258221 and by the Crown center of photonics is gratefully acknowledged. OS is supported by the Adams Fellowship Program of the Israel Academy of Sciences and Humanities. DO is the incumbent of the Recanati career development chair in energy research.


\begin{thebibliography}{30}
\expandafter\ifx\csname natexlab\endcsname\relax\def\natexlab#1{#1}\fi
\expandafter\ifx\csname bibnamefont\endcsname\relax
  \def\bibnamefont#1{#1}\fi
\expandafter\ifx\csname bibfnamefont\endcsname\relax
  \def\bibfnamefont#1{#1}\fi
\expandafter\ifx\csname citenamefont\endcsname\relax
  \def\citenamefont#1{#1}\fi
\expandafter\ifx\csname url\endcsname\relax
  \def\url#1{\texttt{#1}}\fi
\expandafter\ifx\csname urlprefix\endcsname\relax\def\urlprefix{URL }\fi
\providecommand{\bibinfo}[2]{#2}
\providecommand{\eprint}[2][]{\url{#2}}

\bibitem[{\citenamefont{Abbe}(1873)}]{Abbe1873}
\bibinfo{author}{\bibfnamefont{E.}~\bibnamefont{Abbe}}, \bibinfo{journal}{Arch.
  Mikrosk. Anat.} \textbf{\bibinfo{volume}{9}}, \bibinfo{pages}{413}
  (\bibinfo{year}{1873}).

\bibitem[{\citenamefont{Hell and Wichmann}(1994)}]{STED_Hell_OL1994}
\bibinfo{author}{\bibfnamefont{S.}~\bibnamefont{Hell}} \bibnamefont{and}
  \bibinfo{author}{\bibfnamefont{J.}~\bibnamefont{Wichmann}},
  \bibinfo{journal}{Opt. Lett.} \textbf{\bibinfo{volume}{19}},
  \bibinfo{pages}{780} (\bibinfo{year}{1994}).

\bibitem[{\citenamefont{Stefan W.~Hell and Jakobs}(2004)}]{Hell_RESOLFT_2004}
\bibinfo{author}{\bibfnamefont{M.~D.} \bibnamefont{Stefan W.~Hell}}
  \bibnamefont{and} \bibinfo{author}{\bibfnamefont{S.}~\bibnamefont{Jakobs}},
  \bibinfo{journal}{Current Opinion in Neurobiology}
  \textbf{\bibinfo{volume}{14}}, \bibinfo{pages}{599} (\bibinfo{year}{2004}).

\bibitem[{\citenamefont{{Gustafsson}}(2005)}]{Gustafsson_SSI_PNAS05}
\bibinfo{author}{\bibfnamefont{M.~G.~L.} \bibnamefont{{Gustafsson}}},
  \bibinfo{journal}{Proc. Nat. Acad. Sci.} \textbf{\bibinfo{volume}{102}},
  \bibinfo{pages}{13081} (\bibinfo{year}{2005}).

\bibitem[{\citenamefont{Betzig et~al.}(2006)\citenamefont{Betzig, Patterson,
  Sougrat, Lindwasser, Olenych, Bonifacino, Davidson, Lippincott-Schwartz, and
  Hess}}]{PALM_Betzig_Science2006}
\bibinfo{author}{\bibfnamefont{E.}~\bibnamefont{Betzig}},
  \bibinfo{author}{\bibfnamefont{G.}~\bibnamefont{Patterson}},
  \bibinfo{author}{\bibfnamefont{R.}~\bibnamefont{Sougrat}},
  \bibinfo{author}{\bibfnamefont{O.}~\bibnamefont{Lindwasser}},
  \bibinfo{author}{\bibfnamefont{S.}~\bibnamefont{Olenych}},
  \bibinfo{author}{\bibfnamefont{J.}~\bibnamefont{Bonifacino}},
  \bibinfo{author}{\bibfnamefont{M.}~\bibnamefont{Davidson}},
  \bibinfo{author}{\bibfnamefont{J.}~\bibnamefont{Lippincott-Schwartz}},
  \bibnamefont{and} \bibinfo{author}{\bibfnamefont{H.}~\bibnamefont{Hess}},
  \bibinfo{journal}{Science} \textbf{\bibinfo{volume}{313}},
  \bibinfo{pages}{1642} (\bibinfo{year}{2006}).

\bibitem[{\citenamefont{Michael J.~Rust and Zhuang}(2006)}]{STORM_NatMet06}
\bibinfo{author}{\bibfnamefont{M.~B.} \bibnamefont{Michael J.~Rust}}
  \bibnamefont{and} \bibinfo{author}{\bibfnamefont{X.}~\bibnamefont{Zhuang}},
  \bibinfo{journal}{Nature Methods} \textbf{\bibinfo{volume}{3}},
  \bibinfo{pages}{793} (\bibinfo{year}{2006}).

\bibitem[{\citenamefont{Lidke et~al.}(2005)\citenamefont{Lidke, Rieger, Jovin,
  and Heintzmann}}]{Localization_by_blinking_Heintzmann_OpEx2005}
\bibinfo{author}{\bibfnamefont{K.}~\bibnamefont{Lidke}},
  \bibinfo{author}{\bibfnamefont{B.}~\bibnamefont{Rieger}},
  \bibinfo{author}{\bibfnamefont{T.}~\bibnamefont{Jovin}}, \bibnamefont{and}
  \bibinfo{author}{\bibfnamefont{R.}~\bibnamefont{Heintzmann}},
  \bibinfo{journal}{Opt. Express} \textbf{\bibinfo{volume}{13}},
  \bibinfo{pages}{7052} (\bibinfo{year}{2005}).

\bibitem[{\citenamefont{Dertinger et~al.}(2009)\citenamefont{Dertinger, Colyer,
  Iyer, Weiss, and Enderlein}}]{SOFI_DertingerPNAS2009}
\bibinfo{author}{\bibfnamefont{T.}~\bibnamefont{Dertinger}},
  \bibinfo{author}{\bibfnamefont{R.}~\bibnamefont{Colyer}},
  \bibinfo{author}{\bibfnamefont{G.}~\bibnamefont{Iyer}},
  \bibinfo{author}{\bibfnamefont{S.}~\bibnamefont{Weiss}}, \bibnamefont{and}
  \bibinfo{author}{\bibfnamefont{J.}~\bibnamefont{Enderlein}},
  \bibinfo{journal}{Proc. Nat. Acad. Sci.} \textbf{\bibinfo{volume}{106}},
  \bibinfo{pages}{22287} (\bibinfo{year}{2009}).

\bibitem[{\citenamefont{Walther et~al.}(2004)\citenamefont{Walther, Pan,
  Aspelmeyer, Ursin, Gasparoni, and
  Zeilinger}}]{Four_photon_Interference_Zeilinger_Nature2004}
\bibinfo{author}{\bibfnamefont{P.}~\bibnamefont{Walther}},
  \bibinfo{author}{\bibfnamefont{J.}~\bibnamefont{Pan}},
  \bibinfo{author}{\bibfnamefont{M.}~\bibnamefont{Aspelmeyer}},
  \bibinfo{author}{\bibfnamefont{R.}~\bibnamefont{Ursin}},
  \bibinfo{author}{\bibfnamefont{S.}~\bibnamefont{Gasparoni}},
  \bibnamefont{and}
  \bibinfo{author}{\bibfnamefont{A.}~\bibnamefont{Zeilinger}},
  \bibinfo{journal}{Nature} \textbf{\bibinfo{volume}{429}},
  \bibinfo{pages}{158} (\bibinfo{year}{2004}).

\bibitem[{\citenamefont{Afek et~al.}(2010)\citenamefont{Afek, Ambar, and
  Silberberg}}]{Afek_Science2010}
\bibinfo{author}{\bibfnamefont{I.}~\bibnamefont{Afek}},
  \bibinfo{author}{\bibfnamefont{O.}~\bibnamefont{Ambar}}, \bibnamefont{and}
  \bibinfo{author}{\bibfnamefont{Y.}~\bibnamefont{Silberberg}},
  \bibinfo{journal}{Science} \textbf{\bibinfo{volume}{328}},
  \bibinfo{pages}{879} (\bibinfo{year}{2010}).

\bibitem[{\citenamefont{Shin et~al.}(2011)\citenamefont{Shin, Chan, Chang, and
  Boyd}}]{Boyd_CentroidsPRL2011}
\bibinfo{author}{\bibfnamefont{H.}~\bibnamefont{Shin}},
  \bibinfo{author}{\bibfnamefont{K.~W.~C.}~\bibnamefont{Chan}},
  \bibinfo{author}{\bibfnamefont{H.}~\bibnamefont{Chang}}, \bibnamefont{and}
  \bibinfo{author}{\bibfnamefont{R.~W.}~\bibnamefont{Boyd}},
  \bibinfo{journal}{Phys. Rev. Lett.} \textbf{\bibinfo{volume}{107}},
  \bibinfo{pages}{083603} (\bibinfo{year}{2011}).

\bibitem[{\citenamefont{Kolobov and
  Fabre}(2000)}]{Kolobov_Quantum_Limits_PRL2000}
\bibinfo{author}{\bibfnamefont{M.~I.} \bibnamefont{Kolobov}} \bibnamefont{and}
  \bibinfo{author}{\bibfnamefont{C.}~\bibnamefont{Fabre}},
  \bibinfo{journal}{Phys. Rev. Lett.} \textbf{\bibinfo{volume}{85}},
  \bibinfo{pages}{3789} (\bibinfo{year}{2000}).

\bibitem[{\citenamefont{Thiel et~al.}(2007)\citenamefont{Thiel, Bastin, Martin,
  Solano, von~Zanthier, and
  Agarwal}}]{Zanthier_Quantum_imaging_Incoherent_PRL2007}
\bibinfo{author}{\bibfnamefont{C.}~\bibnamefont{Thiel}},
  \bibinfo{author}{\bibfnamefont{T.}~\bibnamefont{Bastin}},
  \bibinfo{author}{\bibfnamefont{J.}~\bibnamefont{Martin}},
  \bibinfo{author}{\bibfnamefont{E.}~\bibnamefont{Solano}},
  \bibinfo{author}{\bibfnamefont{J.}~\bibnamefont{von~Zanthier}},
  \bibnamefont{and} \bibinfo{author}{\bibfnamefont{G.~S.}~\bibnamefont{Agarwal}},
  \bibinfo{journal}{Phys. Rev. Lett.} \textbf{\bibinfo{volume}{99}},
  \bibinfo{pages}{133603} (\bibinfo{year}{2007}).

\bibitem[{\citenamefont{Saleh et~al.}(2005)\citenamefont{Saleh, Teich, and
  Sergienko}}]{saleh2005wolf}
\bibinfo{author}{\bibfnamefont{B.~E.~A.}~\bibnamefont{Saleh}},
  \bibinfo{author}{\bibfnamefont{M.~C.}~\bibnamefont{Teich}}, \bibnamefont{and}
  \bibinfo{author}{\bibfnamefont{A.~V.}~\bibnamefont{Sergienko}},
  \bibinfo{journal}{Phys. Rev. Lett.} \textbf{\bibinfo{volume}{94}},
  \bibinfo{pages}{223601} (\bibinfo{year}{2005}).

\bibitem[{\citenamefont{Tsang}(2009)}]{Centroids_PRL2009}
\bibinfo{author}{\bibfnamefont{M.}~\bibnamefont{Tsang}},
  \bibinfo{journal}{Phys. Rev. Lett.} \textbf{\bibinfo{volume}{102}},
  \bibinfo{pages}{253601} (\bibinfo{year}{2009}).

\bibitem[{\citenamefont{Giovannetti et~al.}(2009)\citenamefont{Giovannetti,
  Lloyd, Maccone, and Shapiro}}]{Shapiro_Sub_Rayleigh_PRA2009}
\bibinfo{author}{\bibfnamefont{V.}~\bibnamefont{Giovannetti}},
  \bibinfo{author}{\bibfnamefont{S.}~\bibnamefont{Lloyd}},
  \bibinfo{author}{\bibfnamefont{L.}~\bibnamefont{Maccone}}, \bibnamefont{and}
  \bibinfo{author}{\bibfnamefont{J.~H.}~\bibnamefont{Shapiro}},
  \bibinfo{journal}{Phys. Rev. A} \textbf{\bibinfo{volume}{79}},
  \bibinfo{pages}{013827} (\bibinfo{year}{2009}).

\bibitem[{\citenamefont{Boyer et~al.}(2008)\citenamefont{Boyer, Marino, Pooser,
  and Lett}}]{Entangled_Images_Science2008}
\bibinfo{author}{\bibfnamefont{V.}~\bibnamefont{Boyer}},
  \bibinfo{author}{\bibfnamefont{A.~M.} \bibnamefont{Marino}},
  \bibinfo{author}{\bibfnamefont{R.~C.} \bibnamefont{Pooser}},
  \bibnamefont{and} \bibinfo{author}{\bibfnamefont{P.~D.} \bibnamefont{Lett}},
  \bibinfo{journal}{Science} \textbf{\bibinfo{volume}{321}},
  \bibinfo{pages}{544} (\bibinfo{year}{2008}).

\bibitem[{\citenamefont{Brida et~al.}(2010)\citenamefont{Brida, Genovese, and
  Berchera}}]{Sub_shot_noise_imaging_NatPhot2010}
\bibinfo{author}{\bibfnamefont{G.}~\bibnamefont{Brida}},
  \bibinfo{author}{\bibfnamefont{M.}~\bibnamefont{Genovese}}, \bibnamefont{and}
  \bibinfo{author}{\bibfnamefont{I.}~\bibnamefont{Berchera}},
  \bibinfo{journal}{Nature Photonics} \textbf{\bibinfo{volume}{4}},
  \bibinfo{pages}{227} (\bibinfo{year}{2010}).

\bibitem[{\citenamefont{Nasr et~al.}(2003)\citenamefont{Nasr, Saleh, Sergienko,
  and Teich}}]{QuantumOCT_PRL2003}
\bibinfo{author}{\bibfnamefont{M.~B.} \bibnamefont{Nasr}},
  \bibinfo{author}{\bibfnamefont{B.~E.~A.} \bibnamefont{Saleh}},
  \bibinfo{author}{\bibfnamefont{A.~V.} \bibnamefont{Sergienko}},
  \bibnamefont{and} \bibinfo{author}{\bibfnamefont{M.~C.} \bibnamefont{Teich}},
  \bibinfo{journal}{Phys. Rev. Lett.} \textbf{\bibinfo{volume}{91}},
  \bibinfo{pages}{083601} (\bibinfo{year}{2003}).

\bibitem[{\citenamefont{Hell et~al.}(1995)\citenamefont{Hell, Soukka, and
  H{\"a}nninen}}]{Hell_photon_pairs1995Bioimaging}
\bibinfo{author}{\bibfnamefont{S.}~\bibnamefont{Hell}},
  \bibinfo{author}{\bibfnamefont{J.}~\bibnamefont{Soukka}}, \bibnamefont{and}
  \bibinfo{author}{\bibfnamefont{P.}~\bibnamefont{H{\"a}nninen}},
  \bibinfo{journal}{Bioimaging} \textbf{\bibinfo{volume}{3}},
  \bibinfo{pages}{64} (\bibinfo{year}{1995}).

\bibitem[{\citenamefont{Kimble et~al.}(1977)\citenamefont{Kimble, Dagenais, and
  Mandel}}]{Antibunching_Kimble_PRL1977}
\bibinfo{author}{\bibfnamefont{H.~J.} \bibnamefont{Kimble}},
  \bibinfo{author}{\bibfnamefont{M.}~\bibnamefont{Dagenais}}, \bibnamefont{and}
  \bibinfo{author}{\bibfnamefont{L.}~\bibnamefont{Mandel}},
  \bibinfo{journal}{Phys. Rev. Lett.} \textbf{\bibinfo{volume}{39}},
  \bibinfo{pages}{691} (\bibinfo{year}{1977}).

\bibitem[{\citenamefont{Basch{\'e} et~al.}(1992)\citenamefont{Basch{\'e},
  Moerner, Orrit, and Talon}}]{Antibunching_dye_Basche_PRL1992}
\bibinfo{author}{\bibfnamefont{T.}~\bibnamefont{Basch{\'e}}},
  \bibinfo{author}{\bibfnamefont{W.~E.}~\bibnamefont{Moerner}},
  \bibinfo{author}{\bibfnamefont{M.}~\bibnamefont{Orrit}}, \bibnamefont{and}
  \bibinfo{author}{\bibfnamefont{H.}~\bibnamefont{Talon}},
  \bibinfo{journal}{Phys. Rev. Lett.} \textbf{\bibinfo{volume}{69}},
  \bibinfo{pages}{1516} (\bibinfo{year}{1992}), ISSN \bibinfo{issn}{1079-7114}.

\bibitem[{\citenamefont{Lounis et~al.}(2000)\citenamefont{Lounis, Bechtel,
  Gerion, Alivisatos, and Moerner}}]{Antibunching_Alivisatos_CPL2000}
\bibinfo{author}{\bibfnamefont{B.}~\bibnamefont{Lounis}},
  \bibinfo{author}{\bibfnamefont{H.}~\bibnamefont{Bechtel}},
  \bibinfo{author}{\bibfnamefont{D.}~\bibnamefont{Gerion}},
  \bibinfo{author}{\bibfnamefont{P.}~\bibnamefont{Alivisatos}},
  \bibnamefont{and} \bibinfo{author}{\bibfnamefont{W.}~\bibnamefont{Moerner}},
  \bibinfo{journal}{Chem. Phys. Lett.} \textbf{\bibinfo{volume}{329}},
  \bibinfo{pages}{399} (\bibinfo{year}{2000}).

\bibitem[{\citenamefont{Michler et~al.}(2000)\citenamefont{Michler, Imamoglu,
  Mason, Carson, Strouse, and Buratto}}]{Antibunching_QDs_Buratto_Nature2000}
\bibinfo{author}{\bibfnamefont{P.}~\bibnamefont{Michler}},
  \bibinfo{author}{\bibfnamefont{A.}~\bibnamefont{Imamoglu}},
  \bibinfo{author}{\bibfnamefont{M.}~\bibnamefont{Mason}},
  \bibinfo{author}{\bibfnamefont{P.}~\bibnamefont{Carson}},
  \bibinfo{author}{\bibfnamefont{G.}~\bibnamefont{Strouse}}, \bibnamefont{and}
  \bibinfo{author}{\bibfnamefont{S.}~\bibnamefont{Buratto}},
  \bibinfo{journal}{Nature} \textbf{\bibinfo{volume}{406}},
  \bibinfo{pages}{968} (\bibinfo{year}{2000}).

\bibitem[{\citenamefont{Mandel}(1979)}]{Sub-Poissonian_Mandel_OL1979}
\bibinfo{author}{\bibfnamefont{L.}~\bibnamefont{Mandel}},
  \bibinfo{journal}{Opt. Lett.} \textbf{\bibinfo{volume}{4}},
  \bibinfo{pages}{205} (\bibinfo{year}{1979}).

\bibitem[{\citenamefont{Schwartz and Oron}(2012)}]{schwartz2012improved}
\bibinfo{author}{\bibfnamefont{O.}~\bibnamefont{Schwartz}} \bibnamefont{and}
  \bibinfo{author}{\bibfnamefont{D.}~\bibnamefont{Oron}},
  \bibinfo{journal}{Phys. Rev. A} \textbf{\bibinfo{volume}{85}},
  \bibinfo{pages}{033812} (\bibinfo{year}{2012}).

\bibitem[{\citenamefont{Michalet et~al.}(2005)\citenamefont{Michalet, Pinaud,
  Bentolila, Tsay, Doose, Li, Sundaresan, Wu, Gambhir, and
  Weiss}}]{QDs_Michalet_Weiss_Science2005}
\bibinfo{author}{\bibfnamefont{X.}~\bibnamefont{Michalet}},
  \bibinfo{author}{\bibfnamefont{F.}~\bibnamefont{Pinaud}},
  \bibinfo{author}{\bibfnamefont{L.}~\bibnamefont{Bentolila}},
  \bibinfo{author}{\bibfnamefont{J.}~\bibnamefont{Tsay}},
  \bibinfo{author}{\bibfnamefont{S.}~\bibnamefont{Doose}},
  \bibinfo{author}{\bibfnamefont{J.}~\bibnamefont{Li}},
  \bibinfo{author}{\bibfnamefont{G.}~\bibnamefont{Sundaresan}},
  \bibinfo{author}{\bibfnamefont{A.}~\bibnamefont{Wu}},
  \bibinfo{author}{\bibfnamefont{S.}~\bibnamefont{Gambhir}}, \bibnamefont{and}
  \bibinfo{author}{\bibfnamefont{S.}~\bibnamefont{Weiss}},
  \bibinfo{journal}{Science} \textbf{\bibinfo{volume}{307}},
  \bibinfo{pages}{538} (\bibinfo{year}{2005}).

\bibitem[{\citenamefont{Schwartz et~al.}(2012)\citenamefont{Schwartz, Tenne,
  Levitt, Deutsch, Itzhakov, and Oron}}]{schwartz2012colloidal}
\bibinfo{author}{\bibfnamefont{O.}~\bibnamefont{Schwartz}},
  \bibinfo{author}{\bibfnamefont{R.}~\bibnamefont{Tenne}},
  \bibinfo{author}{\bibfnamefont{J.}~\bibnamefont{Levitt}},
  \bibinfo{author}{\bibfnamefont{Z.}~\bibnamefont{Deutsch}},
  \bibinfo{author}{\bibfnamefont{S.}~\bibnamefont{Itzhakov}}, \bibnamefont{and}
  \bibinfo{author}{\bibfnamefont{D.}~\bibnamefont{Oron}}, \bibinfo{journal}{ACS
  nano}  (\bibinfo{year}{2012}).

\bibitem[{\citenamefont{Walls and
  Zoller}(1981)}]{Reduced_Fluctuations_Zoller_PRL1981}
\bibinfo{author}{\bibfnamefont{D.~F.} \bibnamefont{Walls}} \bibnamefont{and}
  \bibinfo{author}{\bibfnamefont{P.}~\bibnamefont{Zoller}},
  \bibinfo{journal}{Phys. Rev. Lett.} \textbf{\bibinfo{volume}{47}},
  \bibinfo{pages}{709} (\bibinfo{year}{1981}).

\bibitem[{\citenamefont{Cichos et~al.}(2007)\citenamefont{Cichos, von
  Borczyskowski, and Orrit}}]{Power_law_intermittency_Orrit2007}
\bibinfo{author}{\bibfnamefont{F.}~\bibnamefont{Cichos}},
  \bibinfo{author}{\bibfnamefont{C.}~\bibnamefont{von Borczyskowski}},
  \bibnamefont{and} \bibinfo{author}{\bibfnamefont{M.}~\bibnamefont{Orrit}},
  \bibinfo{journal}{Current Opinion in Colloid \& Interface Science}
  \textbf{\bibinfo{volume}{12}}, \bibinfo{pages}{272 } (\bibinfo{year}{2007}).

\end{thebibliography}

\end{document}